\documentclass{iau}

\usepackage{amsmath}
\usepackage{graphicx}
\usepackage{multirow}

\begin{document}

\lefttitle{Ciprini \& Giacchino}
\righttitle{Unexpected Gamma-ray Outburst from the CSS Radio Galaxy 3C 216}

\jnlPage{1}{7}
\jnlDoiYr{2026}
\doival{10.1017/xxxxx}
\aopheadtitle{Proceedings IAU Symposium No. 406}
\editors{Hanin Kuncarayakti, Tuomas Kangas, \& Melina Bersten eds. \\}

\title{Fermi-LAT Discovery of an Unexpected Gamma-Ray Outburst from the Peculiar Compact Steep-Spectrum Radio Galaxy 3C 216}

\author{Stefano Ciprini$^{1,2}$ and Federica Giacchino$^{3,1}$} 
\affiliation{
$^1$Istituto Nazionale di Fisica Nucleare, INFN, Roma Tor Vergata, I-00133, Italy\\
$^2$Space Science Data Center, SSDC, Agenzia Spaziale Italiana, ASI, Rome, I-00133, Italy\\
$^3$Department of Fundamental Physics, University of Salamanca, Salamanca, E-37008, Spain\\
(on behalf of the \textit{Fermi}-LAT collaboration)\\
}

\begin{abstract}
While compact steep-spectrum (CSS) radio galaxies, genuine young/intermittent-jet or frustrated-jet radio galaxies trapped in gas-rich environments, are rarely detected in GeV gamma-ray energy bands, 3C 216 is an interesting exception. In 2023, the Large Area Telescope (LAT) on board the Fermi Gamma-ray Space Telescope detected an unexpected, rapid, and powerful GeV gamma-ray outburst from this CSS radio galaxy, which was followed up by UV and X-ray observations from the Neil Gehrels Swift Space Observatory. During the outburst its spectral energy distribution is dominated by synchrotron and synchrotron self-Compton radiation, produced in the relativistic jet, suggesting a young quasar classification. Radio-band features, such as superluminal motion and a flat-spectrum core component, confirm that 3C 216 harbors a distinct blazar-like core, bent and viewed under a small angle to the line of sight. This is the unique large and energetic event from the source observed in 17 years of Fermi-LAT continuous all-sky and time-domain monitoring.
\end{abstract}


\maketitle

\section{Introduction}

3C 216 (a.k.a. 4FGL J0910.0+4257, $z=0.670$) is a compact steep-spectrum (CSS) radio galaxy. Along with GHz-peaked spectrum (GPS) extragalactic sources, CSS radio galaxies were historically considered to be exclusively ``young'' radio galaxies at the earliest stages of their evolution ($\lesssim 10^4 - 10^5\text{ yr}$). Currently, CSS radio galaxies are generally divided into two main classes: (1) genuinely young radio sources with expanding relativistic jets that have yet to break through the interstellar medium of the host galaxy; (2) ``frustrated'' radio galaxies trapped within dense, gas-rich galactic environments obstructing jet propagation.

3C 216 exhibits low accretion-flow luminosity and weak broad emission lines in general, but shows powerful radio lobes (i.e. a FR-II radio galaxy morphology), featuring a central radio component, an extended halo ($\sim 56\textrm{ kpc}$ size), and a superluminal jet. Furthermore, its radio core shows a significant spatial misalignment with the outer structure, alongside a low-frequency radio spectrum that turns up at a few GHz, indicating a flat-spectrum central radio component \citep{1993A&A...271...65V,1995AJ....110..522T,akujor96,1999ApJ...525L..13P,   2000PASJ...52..983P,lister2001,2006ApJ...651L..17P,2013MNRAS.429.3551A}

3C 216 is placed in the young jet scenario, with a recently launched, or restarted, powerful accretion and relativistic outflow, testified by a single, intense and short (transient) episode of high-energy gamma-ray production discovered in 2023, by the the Large Area Telescope (LAT) on board the \textit{Fermi} Gamma-ray Space Telescope. Based on this discovery 3C 216 can be considered and re-classified also as a young quasar. The radio and optical radiation of 3C 216 is known to be polarized and variable. Radio polarization indicates a rotation measure that extends further from the core, up to a distance of 2 milliarcseconds into the jet \citep{2013MNRAS.429.3551A}.

\section{The discovery of a prominent GeV gamma-ray outburst}

\begin{figure}[bbb!!!]
 \vspace*{-0.2 cm}
\begin{center}
\hskip -0.6cm 
\resizebox{\hsize}{!}{\rotatebox[]{0}{\includegraphics{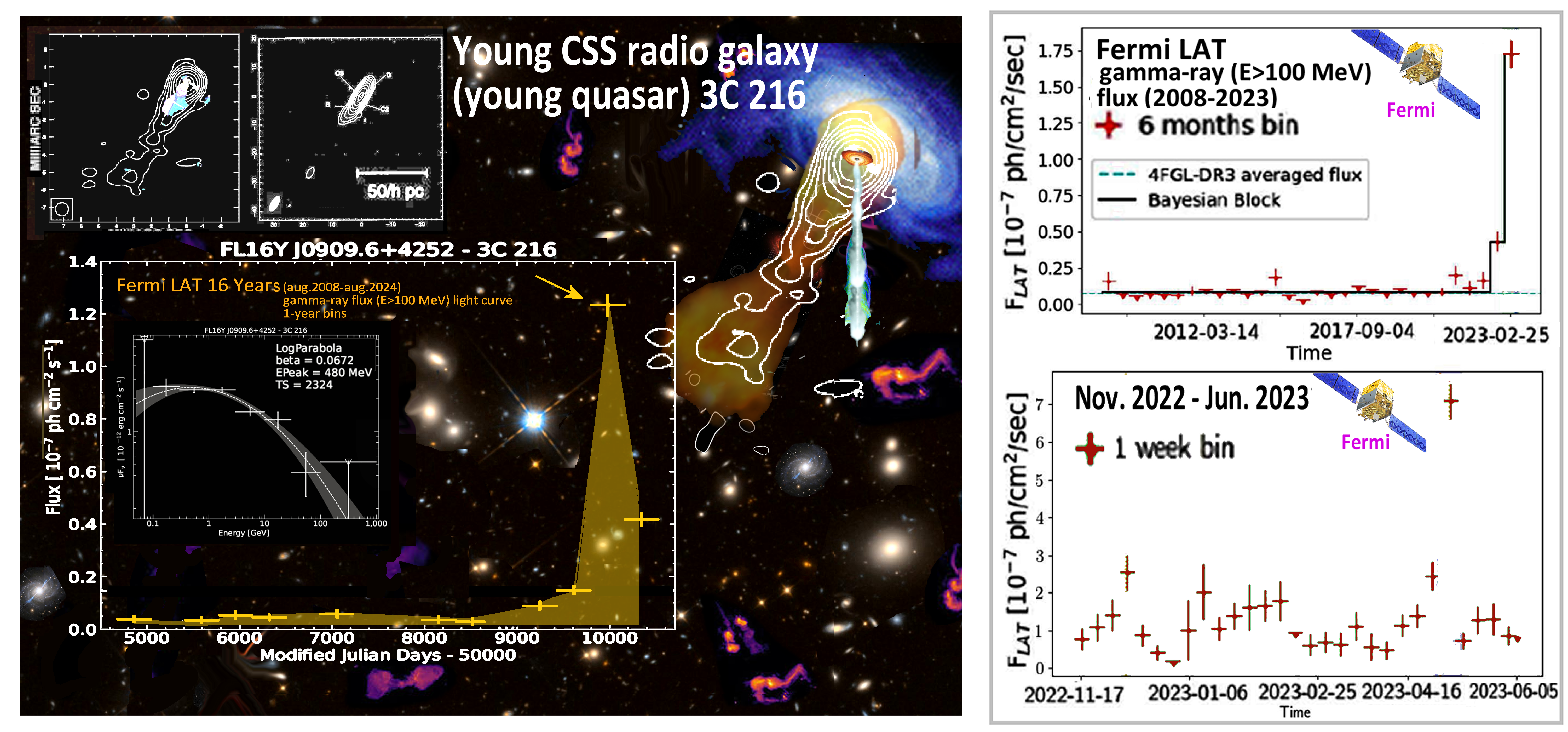}}}
\hskip -0.6cm 
 \vspace*{-0.2 cm}
 \caption{Left panel: pictorial view of 3C 216, its VLBI radio morphology maps, and the 16 year gamma-ray flux light curve, calculated with 1-year bins, provided by the last (early data release) \textit{Fermi}-LAT 16 year Source List (FL16Y) \citep{Ballet2026FL16Y}, with the corresponding curved gamma-ray SED. Right panels: \textit{Fermi}-LAT 100 MeV - 300 GeV flux light curves for the entire (2008 Aug. 04 --- 2023 Jun. 06) epoch, with 6-months time bins, and zoom in the increased activity period (2022 Nov. 14 --- 2023 Jun. 06) with weekly time bins.
}
   \label{fig1}
    \vspace*{-0.7 cm}
\end{center}
\end{figure}

\textit{Fermi}-LAT detected a large-amplitude, rapid, hard GeV gamma-ray outburst from 3C 216 in the period of April and May 2023 (Fig. \ref{fig1}), that represents a unique episode in 17 years of gamma-ray all-sky survey and time-domain monitoring \citep{paper3c216}. A soft-spectrum and low gamma-ray flux were observed in the years prior to this event. The outburst points to a dense population of recently accelerated particles in a region near the central engine, characterized by a relativistically beamed emission viewed at a small angle of sight, resembling a transient blazar-like event/zone, rather than a typical, edge-on CSS radio galaxy.
Mildly enhanced activity has been detected already from about November 2022 (Fig. \ref{fig1}).

%

On May, 5, 2023, 3C 216 showed a sudden outburst, achieving a daily-bin averaged $\gamma$-ray flux of $\langle \Phi \rangle_{daily} = (1.32 \pm 0.15) \times 10^{-6}\, \mathrm{ph\, cm^{-2}\, s^{-1}}$, a factor 176 times higher than the average flux $\langle \Phi \rangle_\gamma = (7.5 \pm 1.4) \times 10^{-9}\, \mathrm{ph\, cm^{-2}\, s^{-1}}$ reported in the 4FGL-DR3 catalog over the $0.1-100\,$GeV energy range. The corresponding photon index decreased (flattened) from the catalog value of $(2.52 \pm 0.10)$ to $(2.11 \pm 0.09)$. This is both the highest daily flux and hardest spectral state reported for 3C 216 to date \citep{2023ATel16024....1G,paper3c216}.

The followed up \textit{Neil Gehrels} Swift observations, joined with \textit{Fermi}-LAT observations pointed out a coherent multifrequency high-flux state from optical-UV, X-ray, to gamma rays (Fig. \ref{fig2}), that can be explained by a single energetic blob with production of energetic radiation, and dominated by the Synchrotron Self-Compton (SSC) emission mechanism. The spectral hardening in gamma rays, is also confirmed by a counterclockwise flux vs photon-index trail, consistent with a single-zone SSC model. The gamma-ray spectrum exhibits significant curvature (Fig. \ref{fig1}), indicating that the radiative efficiency, and the particle energy distribution, are competing between continuous fast energy injection across the entire particle population, and the high-energy radiative cooling.

\section{The Swift ToO follow up}

As X-ray loud AGN, 3C 216 was observed by X-ray space telescopes in the past like Chandra \citep{2022PASJ...74..791K}. A Target of Opportunity (ToO) proposal to the \textit{Neil Gehrels} Swift Observatory, was submitted, for 3C 216, by us and approved (assigned target ID 31845). Swift executed observations of 3C 216 on 2023 May 3, 4, 6, 8, and 9, with visits spaced roughly 1.5 days. XRT operated in Photon Counting (PC) mode, while UVOT carried out exposures using the $V$, $B$, $U$, $UVW1$, $UVM2$, and $UVW2$ photometric filters. X-ray flux was extracted in both the soft ($0.5\text{--}2\text{ keV}$) and hard ($2\text{--}10\text{ keV}$) energy bands. The spectra were fitted using a power-law model modified by neutral Galactic absorption. Over the monitoring period, the flux exhibited a decaying trend in the optical-UV bands, while remaining relatively constant, in a high state, in the X-ray regime (Fig. \ref{fig2}).

\section{Multifrequency SED and modeling}

\begin{figure}[bbb!!!]
 \vspace*{-0.4 cm}
\begin{center}
\hskip -0.6cm 
\resizebox{\hsize}{!}{\rotatebox[]{0}{\includegraphics{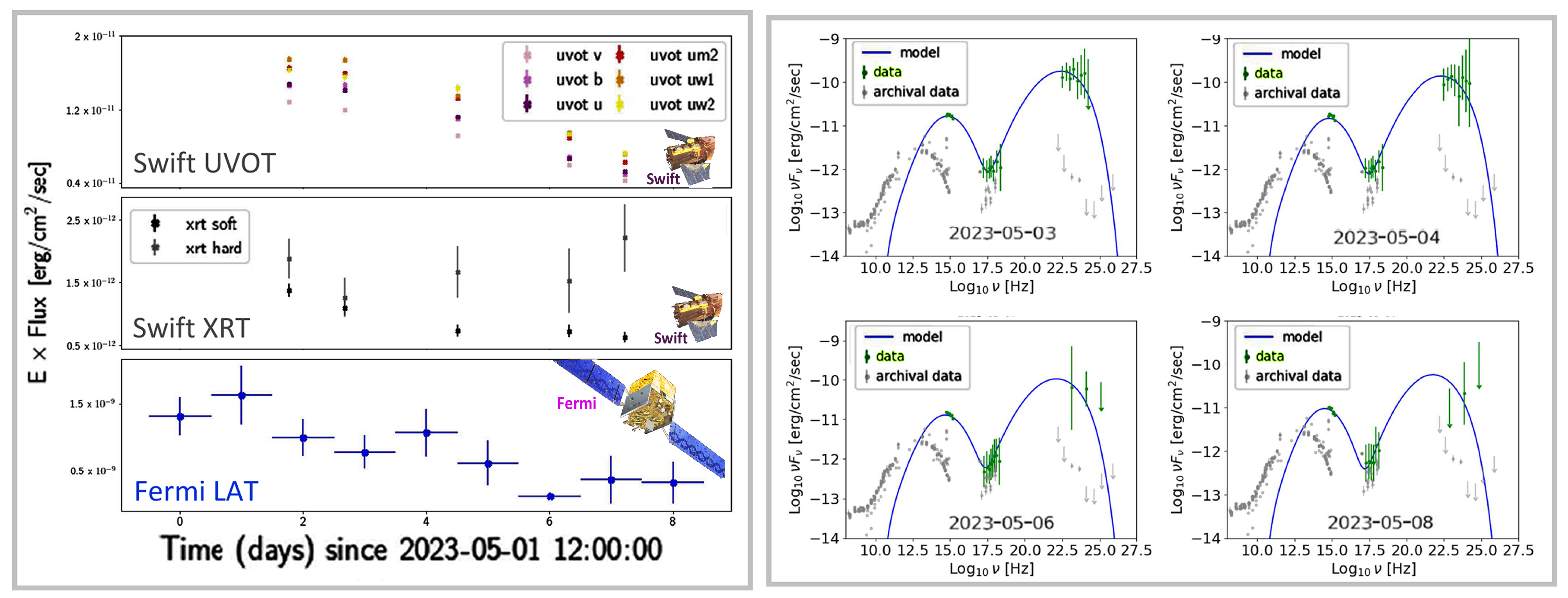}}}
\hskip -0.6cm 
 \vspace*{-0.2 cm}
 \caption{Left panels: Swift UVOT 6 filters (optical and UV) data, and XRT data (two energy ranges) flux light curves, obtained by our ToO observations, and compared to \textit{Fermi}-LAT daily bin gamma-ray ($E>100$ MeV) flux light curve. Right panels: time evolution of the SED for the broadband emission of 3C 216, spanning from optical-UV (Swift XRT and UVOT) to GeV gamma rays (LAT) during the discovered outburst period. Blue line is a single-blob SSC model fit.}
   \label{fig2}
   \vspace*{-0.5 cm}
\end{center}
\end{figure}

Gamma-ray outbursts can be explained by hadronic and leptonic mechanisms, by contributions from external radiation fields and multizone emission scenarios. A SSC model \citep{2008bves.confE..73C} is used to reproduce and fit our simultaneous \textit{Fermi} and Swift data (Fig. \ref{fig2}).
A in-jet shock injects energetic ultra-relativistic electrons, producing strong synchrotron radiation up to the UV and soft X-ray bands. The same population of energetic electrons up-scatter those synchrotron photons to GeV-energy gamma rays, via Inverse Compton scattering. After the outburst, particles cool and synchrotron radiation fades, while Inverse Compton scattering produces hard X-rays rather than gamma rays, as observed in years previous the 2023 \citep{2021MNRAS.507.4564P,paper3c216}. The single-zone SSC model has a lower number of free parameters: ten parameters for a log-parabola particle energy distribution \citep{2020ascl.soft09001T}. Detailed SED modeling results are presented in \citet{paper3c216}.

%

CSS and GPS radio galaxies combine, on subgalactic scales, both the presence of recently injected relativistic particles and abundant photon fields for up-scattering from the central regions of their hosts galaxies.No radio data were collected by us during the rapid outburst, and five simultaneous spectral energy distribution (SED, with \textit{Fermi}-LAT and Swift data) observed during the outburst period, almost day by day, are modeled with SSC (Fig. \ref{fig2}). We provided confirmation of the identification of the gamma-ray source 4FGL J0910.0+4257 with 3C 216. The timescale of approximately 2 days implies a limited size of the active high-energy blob within the relativistic beamed jet responsible for the outburst. The radio jet is misaligned with respect to the radio core and lobes, as found by VLBI data in the past.

\section{Conclusions}

\textit{Fermi}-LAT discovered an extreme gamma-ray outburst from 3C 216, with daily gamma-ray flux surged by a factor of about 176 times above its long-term catalog baseline, with a corresponding hardening of the spectrum, representing, likely, the most violent high-energy outbursts ever observed from a CSS radio galaxy. \textit{Fermi} monitoring and Swift ToO observations pointed out a coherent evolution during the rapid and hard-spectrum outburst, with correlated variability across optical, UV, X-ray, and gamma-ray bands, driven by a in-jet single-blob flaring with Synchrotron Self-Compton mechanism.

3C 216 has, therefore, revealed a blazar-like active zone viewed at a small angle, which is markedly misaligned with the radio lobes. In the literature, 3C 216 was classified as a compact young CSS radio galaxy, as a powerful FR-II radio galaxy and young radio-loud type 2 (torus-obscured) quasar, characterized by extended radio lobes and young relativistic jet, well distinct from classic blazars. Despite its CSS nature, 3C 216 exhibited superluminal radio jet motion ($\sim 4c$), high optical polarization, and a misaligned radio core-jet geometry. This hybrid nature makes its transient gamma-ray behavior, important for understanding AGN physics. Our data and analysis shows that 3C 216 can produce high variability and such a large-amplitude, rapid event from its jet, sharing the same location as blazars in the photon index versus gamma-ray luminosity plot.

Transient high-energy phenomena are energetic, time-variable events where a source undergoes a temporary, drastic increase in luminosity across the electromagnetic spectrum. This gamma-ray outburst from 3C 216 shows that young CSS radio galaxies can generate dramatic accretion events and transient, jet-dominated, blazar-like emission, effectively behaving as transient young quasars on short timescales. Because \textit{Fermi} operates in an all-sky scanning mode, regular gamma-ray monitoring of 3C 216 will continue.

\footnotesize{
\section*{\footnotesize{Acknowledgments}}
The \textit{Fermi}-LAT Collaboration acknowledges support for LAT development, operation and data analysis from NASA and DOE (United States), CEA/Irfu and IN2P3/CNRS (France), ASI and INFN (Italy), MEXT, KEK, and JAXA (Japan), and the K.A.~Wallenberg Foundation, the Swedish Research Council and the National Space Board (Sweden). Science analysis support in the operations phase from INAF (Italy) and CNES (France) is also gratefully acknowledged.
}

\end{document}